\documentclass[usenatbib, useAMS]{mnras}

\usepackage[modulo]{lineno} 
\usepackage{xfrac} 
\usepackage[svgnames]{xcolor}  %
\usepackage{lastpage}  

\newcommand{\arcdeg}{\degr}  
\newcommand{\uv}{\mbox{$u$-$v$}}

\newcommand{\kms}{\mbox{km s$^{-1}$}}

\newcommand{\Jb}{\mbox Jy~beam$^{-1}$}

\newcommand{\thout}{\mbox{$\theta_{\rm o}$}}
\newcommand{\thin}{\mbox{$\theta_{\rm i}$}}

\newcommand{\spost}{_{\rm post} }
\newcommand{\simpact}{_{\rm impact} }
\newcommand{\rx}{\mbox{$r_{\rm 16}$}}
\newcommand{\roi}{\mbox{$\mathcal R_\textnormal{o/i}$}}  

\newcommand{\tablenotemark}[1]{$^{\mathrm #1}$}
\newcommand{\tablenotetext}[2]{\noindent$^{\mathrm #1}$ #2\\}
\newcommand{\phn}{\phantom{1}}

\defcitealias{SN2014C_VLBI}{Paper I}

\title[SN 2014C: VLBI]{SN 2014C: VLBI image shows a shell structure and decelerated expansion}

\author[Bietenholz et al]{Michael F. Bietenholz$^{1,2}$,
Norbert Bartel$^1$,  
Atish Kamble$^3$, \and
Raffaella Margutti$^{4,5}$,
David Jacob Matthews$^{4}$,
and Danny Milisavljevic$^6$
\\
$^1$Department of Physics and Astronomy, York University, Toronto,
M3J~1P3, Ontario, Canada \\
$^2$SARAO/Hartebeesthoek Radio Observatory, PO Box 443, Krugersdorp,
1740, South Africa \\
$^3$formerly at Harvard-Smithsonian Center for Astrophysics, 60
Garden Street, Cambridge, MA 02138, USA \\
$^4$Center for Interdisciplinary Exploration and Research in Astrophysics (CIERA) and Department of Physics and Astronomy,\\
Northwestern University, Evanston, IL 60208, USA \\
$^5$CIFAR Azrieli Global Scholar, Gravity \& the Extreme Universe Program, 2019\\
$^6$Department of Physics and Astronomy, Purdue University, 525 Northwestern Ave., West Lafayette, IN 47907, USA 
}

\begin{document}
\date{Accepted for publication in MNRAS}

\pagerange{\pageref
{firstpage}--\pageref{lastpage}} \pubyear{2021}
\maketitle
\label{firstpage}

\begin{abstract}
  We report on new Very Long Baseline Interferometry radio
  measurements of supernova 2014C in the spiral galaxy NGC~7331, made
  with the European VLBI Network $\sim$5~yr after the explosion, as
  well as on flux density measurements made with the Jansky Very Large
  Array (VLA)\@.  SN~2014C was an unusual supernova, initially of Type
  Ib, but over the course of $\sim$1~yr it developed strong H$\alpha$
  lines, implying the onset of strong interaction with some H-rich
  circumstellar medium (CSM).  The expanding shock-front interacted
  with a dense shell of circumstellar material during the first year,
  but has now emerged from the dense shell and is expanding into the
  lower density CSM beyond.  Our new VLBI observations show a
  relatively clear shell structure and continued expansion with some
  deceleration, with a suggestion that the deceleration is increasing
  at the latest times. Our multi-frequency VLA observations show a
  relatively flat powerlaw spectrum with
  $S_\nu \propto \nu^{-0.56 \pm 0.03}$, and show no decline in the
  radio luminosity since $t\sim1$~yr.
\end{abstract}


\section{Introduction}
\label{sintro}

Supernova (SN) 2014C was a very unusual supernova, and its progenitor
had complex mass-loss in the time before the explosion.  SN~2014C was
discovered on 2014 January 5 in the nearby early-type spiral galaxy
NGC~7331 by the Lick Observatory Supernova Search \citep{Kim+2014}.
We adopt the updated Cepheid distance of $D = 15.1 \pm 0.7$~Mpc from
\citet{Saha+2006}\footnote{The NASA/IPAC Extragalactic Database (NED),
  {\url{https://ned.ipac.caltech.edu}}, lists 54 redshift-independent
  distances, with mean and standard deviation $13.4 \pm 2.7$~Mpc.},
and an explosion date, $t = 0$, of 2013 December 30.0 (UT) = MJD
56656.0, as determined by \citet{Margutti+2017_SN2014C} from
bolometric lightcurve modelling.\footnote{Although SN~2014C's
  explosion date is not tightly constrained, it is uncertain by less
  than one week, and the exact value will have little effect on our
  results which are at times several years after the explosion.}

At its discovery, SN~2014C had the spectrum of ordinary, H-stripped
Type Ib supernova \citep{Kim+2014,Tartaglia+2014}.  Unfortunately, no
spectra could be obtained for several months thereafter as it went
behind the sun, but after it emerged, the spectrum had evolved into a
Type IIn one, with prominent H$\alpha$-lines, which implied strong
interaction with the circumstellar medium (CSM)
\citep{Milisavljevic+2015_SN2014C, Margutti+2017_SN2014C}.  At
$t = 20$~d, it was faint in X-rays, but, unusually, rose till
$t \sim 1$~yr and has remained high since
\citep{Margutti+2017_SN2014C,JinK2019}.  In the mid-infrared, it had a
high and almost constant brightness till $t \sim 5.5$~yr
\citep{Tinyanont+2019}.

SN 2014C was quickly also detected in the radio, at frequencies
ranging from 7~GHz to 85~GHz \citep{Kamble+2014b, Zauderer+2014}.  In
the first month, it did not have a high radio luminosity
\citep[$L_\nu \sim 10^{26}$~erg~s$^{-1}$~Hz$^{-1}$ at 7
GHz,][]{Kamble+2014b}\@.  However, the luminosity rose rapidly after
about one month, and then again around 1~yr \citep{Anderson+2017},
and has stayed high.  Due to its relative nearness and high radio
brightness it was a target for Very Long Baseline Interferometry
(VLBI) observations.  In \citet{SN2014C_VLBI}, which we will refer to
as \citetalias{SN2014C_VLBI} hereafter, we presented our first four
epochs of VLBI observations, between $t = 1.1$ and 2.9~yr, which
showed that SN~2014C's forward shock was expanding at
$v \simeq 13\,600$~\kms\ over that period, but must have been
expanding more rapidly at $t < 1$~yr.

The picture of SN~2014C that has emerged \citep[e.g.][]
{Milisavljevic+2015_SN2014C, Margutti+2017_SN2014C} is that the
supernova exploded as a Type Ib, having lost most of its H envelope,
inside a low-density cavity.  Due to the low density there was
relatively little radio or X-ray emission initially.  As the shock
moved outward, at $t \sim 0.3$~yr it encountered a shell of very dense
circumstellar material (CSM), causing emission at both X-rays and
radio to brighten.  Interaction with the CSM commonly produces both
radio and X-ray emission \citep[e.g.][]{ChevalierF2017}.

The shell was formed due to a mass ejection event
shortly before the SN explosion.  The shock has since progressed
through this overdense shell, and is currently expanding through the
moderately dense wind of the progenitor from the time before the
ejection event.

In the few cases, like SN~2014C, where a SN is near enough to be
resolved with VLBI observations, they can provide crucial direct
observational constraints on basic physical parameters of the SN, in
particular the (time-dependent) radius of the expanding ejecta and the
corresponding expansion speed.  In \citetalias{SN2014C_VLBI}, we used
VLBI observations to determine the radius, $r$, of the shock in
SN~2014C at various epochs, and found a radius of
$\rx = 14.4 \pm 0.6$, at $t = 2.9$~yr, where \rx\ is a dimensionless
radius, and is equal to $r/(10^{16} \; {\rm cm})$.

In order to continue to study the evolution of SN 2014C, we made new
VLBI observations, this time with the European VLBI Network (EVN),
$\sim$2~yr after those presented in \citetalias{SN2014C_VLBI}.  Our
new image has the highest resolution relative to the shell size for
this SN to date.  We also report on observations with the Karl
G. Jansky Very Large Array of the National Radio Astronomy Observatory
(NRAO) in the U.S.A.\ to measure the total flux density and spectral
energy distribution (SED) in early 2020.

\section{Observations and Data Reduction}
\label{sobs}

\subsection{VLBI Observations}
\label{svlbiobs}

We observed SN~2014C using the EVN on 2018 October 30 and 31 at
8.4~GHz (observing codes EB066A, EB066B).  Both observations used a
standard 1-Gbps experiment setup
(8 sub-bands, 16 MHz bandwidth per sub-band, dual circular
polarisation, 2-bit quantisation). The participating telescopes were
Westerbork (Wb, phased-array), Effelsberg (Ef), Medicina (Mc),
Onsala(O6), Tianma (T6), Urumqi (Ur), Yebes (Ys), Hartebeesthoek (Hh)
Svetloe (Sv), Zelenchukskaya (Zc), Badary (Bd) and Irbene (Ir).
The correlation was done by the EVN software correlator
\citep[SFXC;][]{Keimpema+2015} at JIVE (Joint Institute for VLBI, ERIC)
using standard correlation parameters of continuum experiments.
Each of the two runs was 8~h in length.

\begin{figure}
\centering
\includegraphics[width=0.35\textwidth]{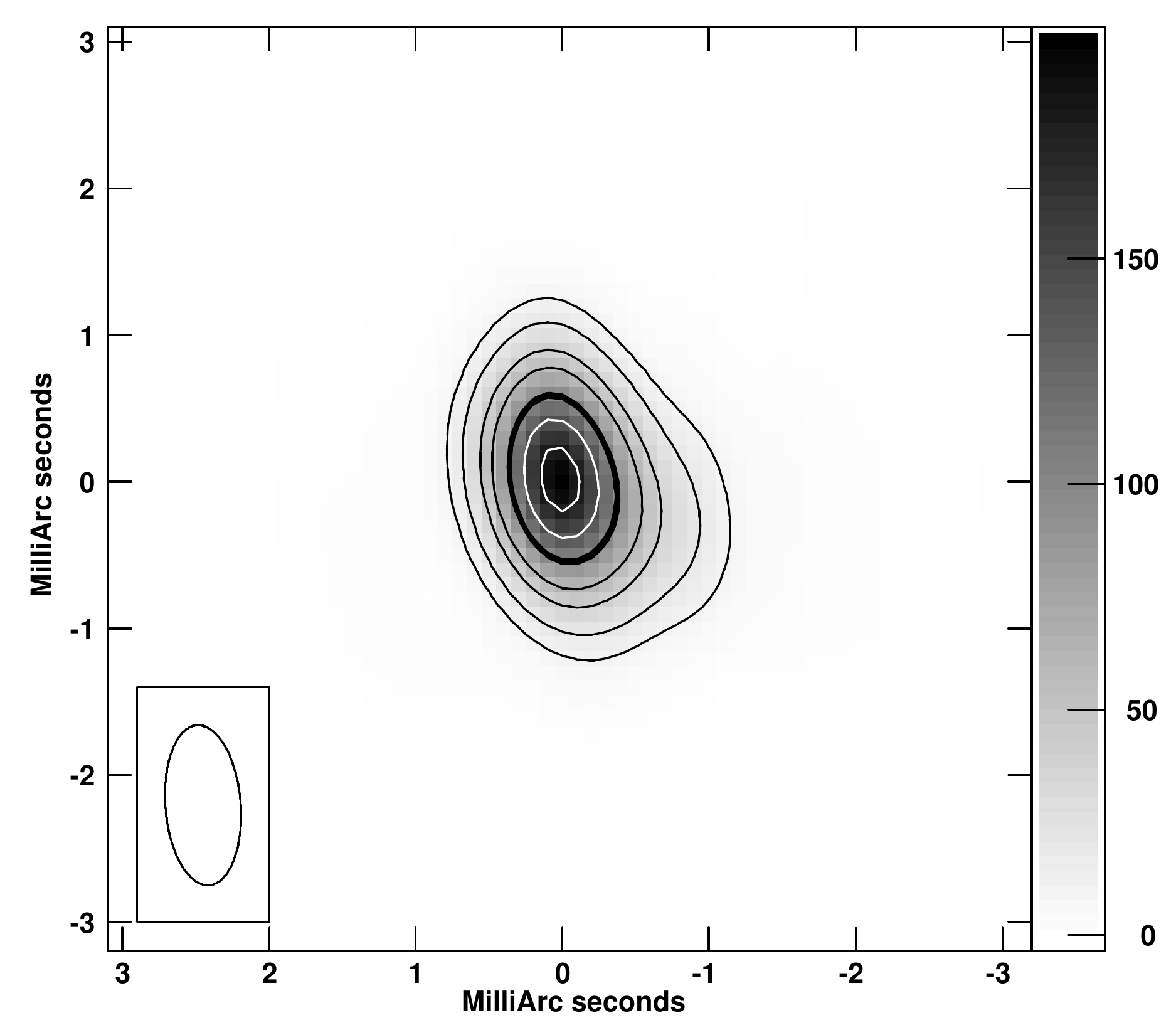}
\caption{The VLBI image our phase-reference source J2248+3718\@.  The
  contours are drawn at 5, 10, 20, 30, {\bf 50} (emphasized), 70 and
  90\% of the peak brightness of 198 m\Jb.  The full width at half
  maximum (FWHM) resolution of ($1.09 \times 0.51$)~mas at p.a.\
  4\arcdeg\ is indicated at lower left. North is up and east is to the
  left.}
\label{f2248}
\end{figure}

We phase-referenced our observations to the source VCS1 J2248+3718,
which we will refer to as just J2248+3718, and which is 2.9\arcdeg\
away from SN~2014C on the sky.  We found it to be only marginally
resolved\footnote{In \citetalias{SN2014C_VLBI} we had used the nearer
  NVSS J223555+341837 as a phase-reference source, but we found it to
  be significantly resolved, and we therefore switched to the
  less-resolved J2248+3718 for these observations.  Since we were able
  to phase self-calibrate SN~2014C at 8.4~GHz both in this work and in
  \citetalias{SN2014C_VLBI}, any structure in reference sources should
  not affect our results.}.
  We show the image of J2248+3718 in Figure~\ref{f2248}.  Our
  phase-referencing calibration for SN~2014C, which provided the
  starting point for the phase self-calibration of SN~2014C, was based
  on the CLEAN model of J2248+3718.

The data reduction was carried out with NRAO's Astronomical Image
Processing System (\textsc{AIPS}).  The initial flux density calibration was
done through measurements of the system temperature at each telescope,
and improved through self-calibration of the phase-reference sources.

The signal-to-noise ratio on SN~2014C was high enough to permit
self-calibration in phase. We started with self-calibrating the
antennas T6, Bd, and Ir, which showed the most obvious failures in
phase-referencing and exhibited phase-wrapping, due in part to
inaccurate antenna positions.  We used a 15~min solution interval, and
an initial CLEAN model made {\em excluding}\/ the data from those
three antennas.  We then proceeded to include those three antennas in
the imaging and the other antennas in the self-calibration, with a
longer solution interval of 2~h, but overlapped so that we obtained a
solution every hour.  Both imaging and model-fitting results for
SN~2014C are derived from the phase self-calibrated data.
  
\subsection{VLA Observations}
\label{svlaobs}

We observed SN~2014C also with the VLA on 2020 May 6, (observing code
20A-441)\@.  The total length of the observing run was 2~h, and we
observed at frequencies between 2 and 20 GHz.  The data were reduced
following standard procedures using the Common Astronomy Software
Application \citep[\textsc{CASA};][]{McMullin+2007}, with the flux
density scale set by observations of 3C~286\@.  The SN~2014C data were
self-calibrated in phase only.

We measured flux densities by fitting an elliptical Gaussian of the
same dimension as the CLEAN beam to the image, with a zero-level also
being fit in cases where there was significant background emission
from the galaxy, although in all cases the galaxy background was less
than the uncertainties.  Our uncertainties include the statistical
contribution due to the noise in the images, but are dominated by the
5\% uncertainty in the flux-density calibration at the VLA.

\section{Results}
\label{sresults}

\subsection{VLBI image}
\label{svlbiimg}

\begin{figure}
\centering
\includegraphics[width=0.45\textwidth]{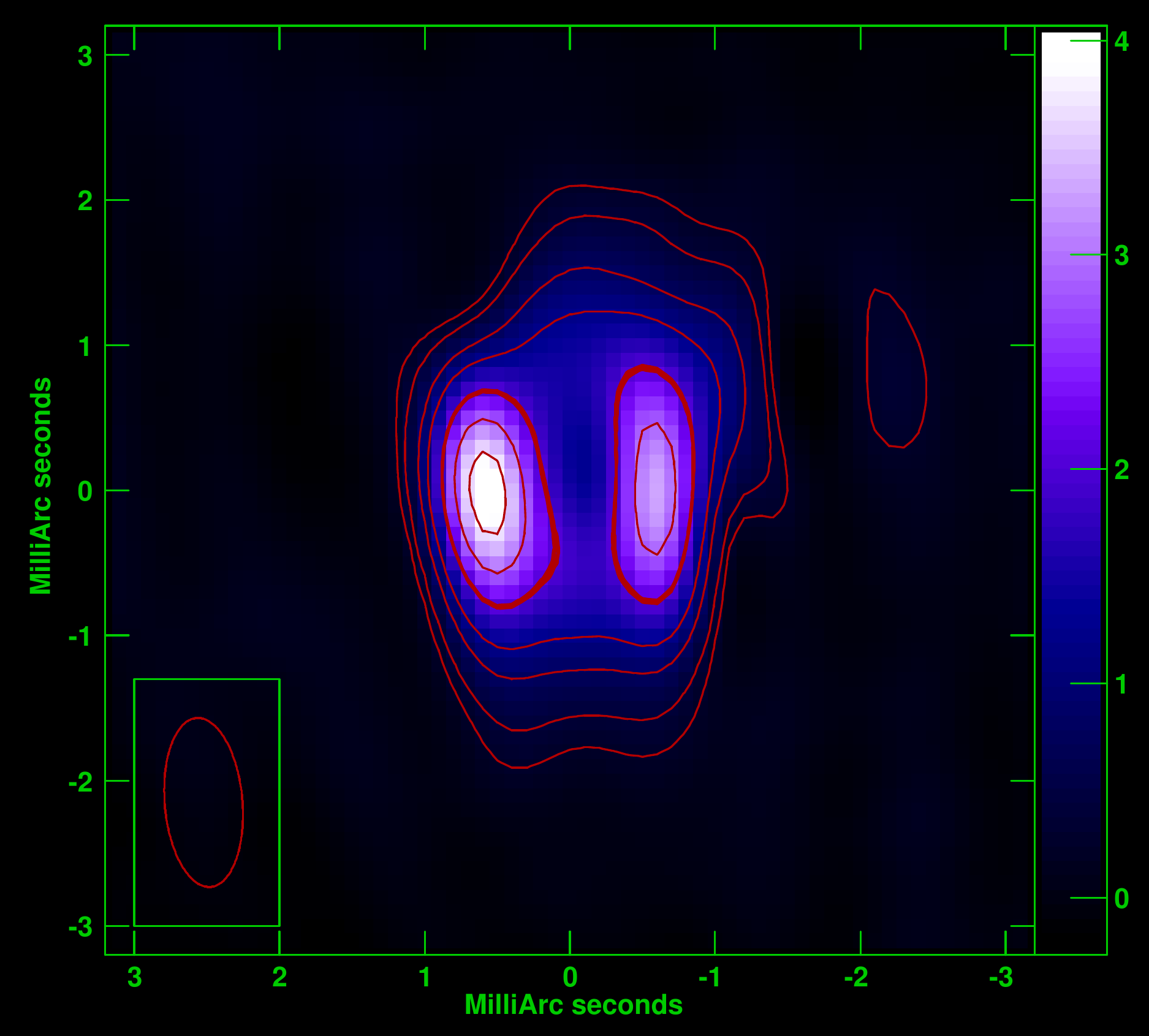}
\caption{The VLBI image of SN~2014C on 2018 Oct.\ 31 at $t = 4.8$~yr.
  Both the contours and colourscale show the brightness, the latter
  labelled in m\Jb.  The contours are at $-6$, 6, 10, 30, {\bf 50}
  (emphasized), 70 and 90\% of the image peak brightness of
  4080~$\mu$\Jb. The rms background brightness was 51~$\mu$\Jb.  The
  FWHM resolution of ($1.17 \times 0.54$)~mas at p.a.\ 5\arcdeg\ is
  indicated at lower left.  North is up and east is to the left.}
\label{fimage}
\end{figure}

We show the VLBI image of SN~2014C, obtained on 2018 Oct.\ 31, or
$t = 4.8$~yr, in Figure~\ref{fimage}.  The image was deconvolved using
the CLEAN algorithm, with AIPS robustness parameter set to +0.5.  To
increase the reliability of the images we used the square root of the
data weights in the imaging, which results in more robust images less
dominated by a small number of very sensitive baselines.  We also use
the multi-scale extension of the original CLEAN algorithm, MS-CLEAN
\citep{WakkerS1988}, which produces superior results for extended
sources \citep[see, e.g.,][]{Hunter+2012, Rich+2008, SNR4449}.  The
total CLEANed flux density was 15.8~mJy, the rms background brightness
was 51~$\mu$\Jb, and the FWHM resolution was $(1.17 \times 0.54)$~mas
at p.a.\ 5\arcdeg.

The image shows a structure that is at least approximately circular in
outline, with enhancement to the east and west, with the one in the
east being about 25\% brighter.
An east-west asymmetry of similar magnitude was seen in our image from
$t = 2.9$~yr in \citetalias{SN2014C_VLBI}, but in the opposite sense,
with the west side being brighter.  Such one-sided asymmetries in the
radio brightness seem to be common in SNe \citep{SNVLBI_Cagliari,
  BartelKB2017}, and they can vary with time \citep[e.g.\ SN
1993J,][]{SN93J-3}, but their origin is not known.

\begin{figure}
\centering
\includegraphics[width=0.36\textwidth]{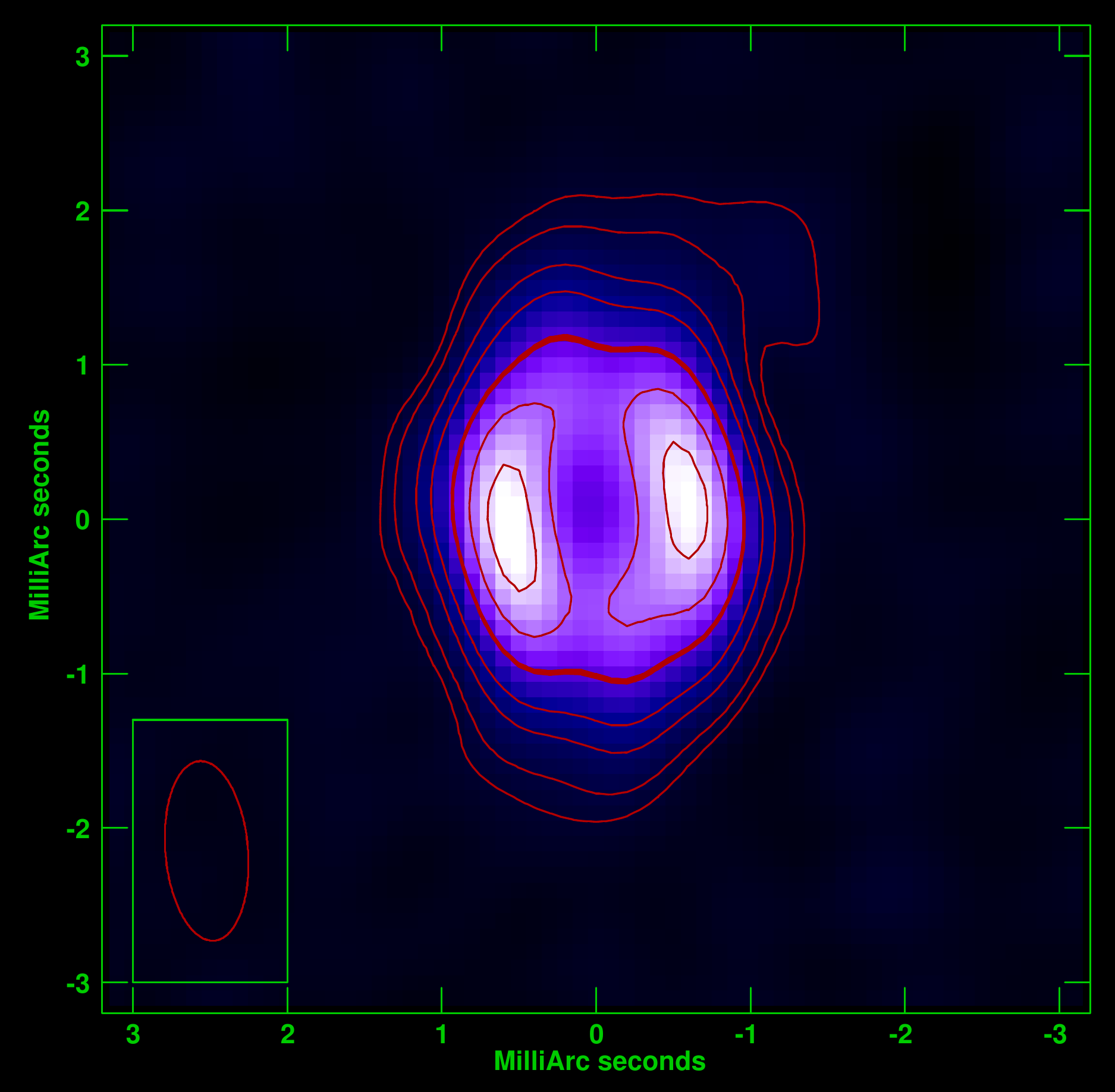}
\caption{An image made by deconvolving model visibility data, where
  the \uv~plane model was a completely spherically symmetrical shell.
  Random noise was added to the model visibilities to match the
  observed image.  The deconvolution was the same as was used
  for the observed image.  The model had outer angular radius, \thout\
  = 0.89~mas and \roi\ = 1.1.  Despite being completely symmetrical,
  when convolved with the elliptical beam, the model image shows two
  hot-spots very similar to those in the observed image
  (Figure~\ref{fimage}).}
\label{fmodel}
\end{figure}

Is the enhancement in brightness to the east and west real, or is it
merely due to the convolution of a circular ring-like brightness
pattern with a north-south elongated restoring beam?  To answer this
question, we simulated visibility measurements for a source with
complete circular symmetry, which simulated visibilities had the same
elongated \uv~coverage, and thus the same elongated restoring beam as
our EVN observations.  We added random Gaussian noise the simulated
visibilities, scaled so as to produce the same image background rms as
was found in the image made from the real data. We then deconvolved
these simulated visibilities in the same fashion as the real data.
Our source model was the projection of a spherical shell of emission,
with a ratio of outer to inner radius, \roi = 1.1 (we justify this
choice in Section \ref{ssize} below). We show the resulting simulated
image in Figure~\ref{fmodel}.  The image looks very similar to the
real VLBI image in Figure~\ref{fimage}, in particular in also having
enhanced brightness to the E and W.

While the real image has brightness contrasts of $\sim$2.2:1 between
the hot spots to the east and west and the corresponding ``gaps'' to
the north and south, the simulated one with \roi\ = 1.1 had brightness
contrasts of 1.4:1, which would increase if smaller values of
\roi\ were used for the model.  This suggests that a significant part
of the brightness enhancement to the east and west in our image is due
merely to our elongated beam, although there may also be some real
enhancement particularly to the east, where the observed image has
higher brightness.

Our simulated data differ from the real measurements in one respect.
Although we scaled the noise added to the simulated visibilities to
produce the same image background rms, the noise we added was
uncorrelated between visibilities.  The real visibilities, on the
other hand, are corrupted both by random and uncorrelated noise and by
residual calibration errors. Since the calibration is antenna-based,
this introduces correlations in the visibility errors for the real
data not present in the simulated data.  Since we phase
self-calibrated the residual calibration errors should be small, and
it seems unlikely that such correlated errors would cause systematic
changes in the apparent image morphology.

\subsection{Size and  Expansion Speed}
\label{ssize}

To determine a precise size for SN~2014C, we fit a geometrical,
spherical-shell model in the Fourier transform or \uv~plane, as we did
in \citetalias{SN2014C_VLBI}\footnote{\citet{SN93J_Manchester} showed
  that in the case of SN~1993J, the results obtained through \uv~plane
  modelfitting are superior to those obtained in the image plane.}.

We used the same model we used in \citetalias{SN2014C_VLBI}, which is
the Fourier transform of the projection of an optically thin shell of
emission.  The model is characterized by the inner and outer angular
radii of the shell, $\thin, \, \thout$, and the total flux density.
We again used the square root of the data weights in the fitting,
which makes the results more robust at the expense of some statistical
efficiency.

We justify this choice of model geometry for SN~2014C in
\citetalias{SN2014C_VLBI}, and the same geometry has been found
appropriate for other SNe \citep[e.g.][]{SN93J-2, SN2011dh_Alet}.  It
is the outer angular radius, \thout, which is most closely identified
with the forward shock, and which is also most reliably determined by
the data.  We therefore first fix the ratio of \roi\ (= \thout/\thin)
to 1.25, which has been shown to be appropriate in the case of
SN~1993J \citep{SN93J-3, SN93J-4}. For the case of a simple CSM
structure and a non-magnetic shell, similar values were also seen in
numerical simulations \citep {JunN1996a}.
The fitted value of \thout\ is only weakly dependent on the assumed
value of \roi.

As we found for earlier epochs, the purely statistical uncertainty on
\thout\ was small, $\sim$0.6\%\@.  We follow the same procedure as in
\citetalias{SN2014C_VLBI} to estimate a systematic uncertainty, and
again include three contributions in our final standard error, added in
quadrature.

The first contribution was estimated using jackknife re-sampling
\citep{McIntosh2016}.  Specifically, we dropped the data from each of
the antennas in the VLBI array in turn and calculated
$N_{\rm antenna} = 12$ new estimates of the fitted size, and the
scatter over these 12 values allows one to estimate the uncertainty of
the original value that included all antennas.  We obtained a
jackknife relative uncertainty of 6.5\%.

The second contribution is an estimate of the effect of any residual
mis-calibration of the antenna amplitude gains on the fitted sizes.
We estimated this contribution to the uncertainty in a Monte-Carlo
fashion by repeatedly randomly varying the individual antenna gains by
10\% (rms), and then re-fitting the spherical shell models.  This
estimate should be conservative as it is unlikely that our antenna
gains would be wrong by as much as 10\%.  We find this causes a 1.4\%
uncertainty in the fitted radius.  The fitted angular outer radius
with the full uncertainty is then $\thout = 0.94 \pm 0.06$~mas, with
the assumption of \roi = 1.25.

The fitted value of \thout\ does depend weakly on \roi\@.  Therefore,
rather than assuming \roi = 1.25 we attempted to fit \roi\ in
addition to \thout\@.  The result suggests a thin shell, with the
best-fit value being \roi\ = 1, and a corresponding best-fit \thout =
0.85~mas, about 0.09 mas or $1.3\sigma$ smaller than the value
obtained with the assumption of \roi\ = 1.25.

The assumption of a completely optically-thin shell is likely not
warranted as the unshocked ejecta are expected to remain optically
thick to radio waves for decades \citep{MioduszewskiDB2001,
  SN86J-FRB}.  This will affect the fitted value of \roi, in the sense
that use of a completely optically-thin model will cause \roi\ to be
overestimated.  This is likely the reason why our best fit value of
\roi\ is near unity, the maximum possible value, and the true value is
likely to be lower.

An uncertainty on \roi\ is difficult to estimate.  By definition,
\roi\ cannot be $<1$, so any error in the estimate cannot be
Gaussian-distributed.  Since our FWHM resolution, even in the more
well resolved east-west direction, is only 0.54 mas, reliably
determining the shell thickness of $\la 0.20$~mas seems a tall
order.  We therefore cannot precisely determine \roi, but we can say
that it is likely between 1 and 1.25, with smaller values,
corresponding to thinner shells, being more probable.

As might be expected, the inferred value of the outer radius, \thout,
is only very weakly dependent on absorption in the centre.  In the
case of SN~1993J, \citet{SN93J-3} fitted more elaborate models which
allowed for absorption in the centre of the SN and found almost no
effect on the fitted values of \thout, so the shortcomings of our
optically-thin model are unlikely to significantly affect our values
of \thout.

We then take our final value for \thout\ to be the midpoint
of the two values obtained for \roi\ fixed at 1.25, and \roi\ free
(with the fitted value \roi\ = 1), and add in quadrature half the
difference in those two values of \thout\ to the uncertainty
we had determined in the fixed \roi\ = 1.25 case, to obtain a final
value for \thout\ of $0.89 \pm 0.08$~mas.  At the distance of
SN~2014C (15.1 Mpc), this radius corresponds to a linear size of
$\rx = 20.1 \pm 1.7$.

\subsection{Expansion Curve}
\label{sexpand}

\begin{figure}
  \centering
\includegraphics[width=\linewidth]{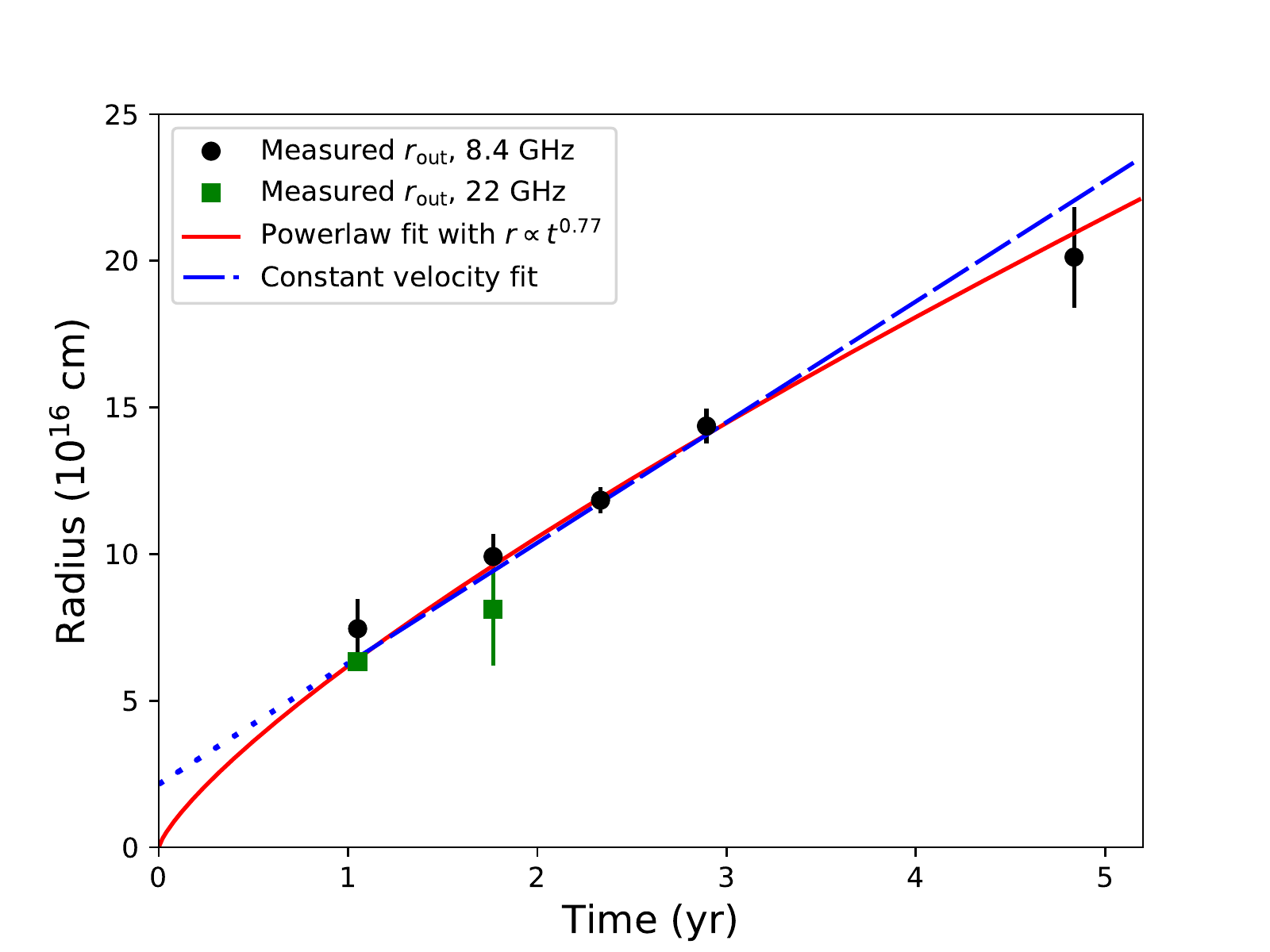}
\caption{The radius of SN~2014C as a function of time, $t$, since the
  explosion at $t=0$ on 2013 Dec.\ 30.  The outer radii were
  determined by fitting a spherical shell model directly to the
  visibilities in this paper and in \citetalias{SN2014C_VLBI}, and
  calculated for a distance of $D = 15.1$~Mpc.  Radii measured at 8.4
  GHz are shown as black circles and those at 22 GHz as green
  squares. We show two different functions fitted to the measured
  radii.  The first, shown by the solid (red) line, is an
  uninterrupted powerlaw expansion of the form $r \propto t^{m=0.77}$.
  The second, shown by the dashed (blue) line, is a constant velocity
  expansion after $t = 1$~yr (with an implied more rapid expansion
  before then).  We expect the approximately constant-velocity regime
  to begin at $t \sim 1$~yr; hence, we show the extrapolation of the
  constant velocity fit to earlier times with a dotted line.}
\label{fexpand}
\end{figure}

We plot our new value for the radius at $t=4.8$~yr along with earlier ones
from \citetalias{SN2014C_VLBI}, in Figure~\ref{fexpand}.  The
expansion of SNe is often parameterized as a powerlaw, such that
$r = r_{\rm 1yr} (t \, /{\rm \, yr})^m$, where $r$ is the radius of
the supernova at time $t$, $r_{\rm 1yr}$ is the radius at $t = 1$~yr,
and $m$ is the powerlaw coefficient, often called the expansion
parameter.  Such a function has been shown to be expected on
theoretical grounds with $m$ in the range 0.6 to 1
\citep[e.g.][]{Chevalier1982b}, and used to describe other SNe
\citep[e.g.\ SN~1993J][]{SN93J-2}.

The velocity between our previous measurement of $r$ at $t$ = 2.9 yr
(Paper I) and the present one at $t = 4.8$~yr is $9400 \pm
2900$~\kms. If we interpret the evolution as a powerlaw, the values at
$t = 2.9$ and 4.8~yr imply $m = 0.66 \pm 0.18$, suggesting that,
compared to the average velocity since the explosion, there is a
moderate amount of deceleration over the last two measurements.

We turn now to fitting all the radius measurements.   We fit the same
two functions we used in \citetalias{SN2014C_VLBI} to our measurements
of \rx\ by least squares, and we refer the reader to that paper for a
fuller discussion of the choice of functions.  The first
function is the powerlaw function just described.

Fitting a powerlaw function to our radius measurements, we obtain:
$$\rx = (6.27 \pm 0.22) \cdot \left(\frac{t}{\rm 1 \,
    yr}\right)^{(0.77 \pm 0.03)} \left(\frac{D}{15.1\,{\rm
      Mpc}}\right),$$ with a sum of squared residuals, SSR = 2.4\@.
We plot this fitted expansion curve as the red line in Figure
\ref{fexpand}.  The fitted value $m$ is higher, albeit not
significantly so, than that of $m=0.66\pm0.18$ obtained from only the
last two measurements, suggesting a possible increase in deceleration
at the latest times.

The fitted expansion curve, with $m = 0.77$, suggests a moderate
amount of deceleration over the history of the SN\@.  This value of
$m$ is consistent with what is generally expected from the mini-shell
model.  If the CSM has a wind density profile ($\rho \propto r^{-2}$),
then the mini-shell solution has that $m = (n - 3)/(n - 2)$
\citep{Chevalier1982b}, so our value of $m$ suggests ejecta with
$\rho \propto r^{-n}$ with a relatively flat value of
$n = 6.5^{+0.8}_{-0.6}$.

In the self-similar solution of \citet{Chevalier1982a}, the value of
\roi\ depends on $n$, and for $n = 6.5$ a value of $\roi \simeq 1.3$
is expected, whereas our model-fitting suggested smaller values of
$\roi \, \la \, 1.25$ (Section~\ref{ssize}).  However, since the
density structure of SN~2014C's CSM was clearly more complex than a
simple $\rho \propto r^{-2}$ powerlaw assumed in the self-similar
model, we should expect that the relationships between $m$, $n$ and
\roi\ will deviate somewhat from those in the self-similar case.

There are, however, good reasons to think that a simple powerlaw may
not be appropriate to describe SN~2014C's expansion.  As we described in the
introduction, at about $t \sim 0.3$~yr, SN~2014C's expanding shock
seems to have encountered a region of dense, H-rich CSM, leading to an
evolution that deviates from the self-similar powerlaw function of the
mini-shell model.  Systems of this nature have been considered by
numerous authors \citep{ChevalierL1989, ChugaiC2006, SmithM2007,
  vMarle+2010}.
  In such a system, the shock slows dramatically when it first
  encounters the dense shell.  It then accelerates as it emerges from
  the dense CSM shell, and subsequently proceeds to coast at almost
  constant speed until the mass of the CSM swept up from outside the
  massive shell becomes comparable to the shell mass, at which point
an approximately power-law expansion resumes.  This behaviour has been
reproduced in numerical simulations by \citet{vMarle+2010}.

The impact of the SN shock on the dense CSM shell for SN 2014C
occurred at $t \sim 0.3$~yr, before the first VLBI observations at
$t = 1.1$~yr.  We cannot, therefore, directly resolve the slowing of
the shock, so we model only the period of approximately
constant-velocity expansion after the impact of the shock on the
massive shell.  Hence, the second function that we fit to SN~2014C's
expansion, which we call the ``constant velocity'' function, is
$r [t > t\simpact] = r\simpact + v\spost \cdot (t - t\simpact)$, where
$t\simpact$ is the time at which the shock impacts on the dense shell,
$r\simpact$ is the radius at that time, and $v\spost$ is the shock
velocity after that time.  For $t\simpact \leq 1$~yr, that function is
equal to $r = r_{\rm 1yr} + v\spost \cdot (t - 1\;{\rm yr})$, so we
fit the latter function and avoid the problem of not knowing
$t\simpact$ exactly.  It is expected that $t\simpact$ is in the range
$0.3 \sim 0.6$~yr \citep[e.g.][]{HarrisN2020}\@.  We again fit the
function to the VLBI radius measurements using weighted least squares.

Note that the powerlaw function also produces constant-velocity
expansion when $m = 1$, but there is a crucial difference between the
two functions: the powerlaw function with $m=1$ is just uninterrupted
free expansion starting from $r$=0, $t$=0, whereas our constant
velocity function only has a constant velocity after $t=1$~yr, and
does not extrapolate to $r$=0, $t$=0, since a more rapid expansion at
$t < 1$~yr is implicit.

Fitting the constant-velocity function, we obtained
$$\rx = (6.27 \pm 0.22) + (4.12 \pm 0.22) \times
 \left(\frac{t}{\rm 1 \,
  yr} - 1\right) \left(\frac{D}{15.1\,{\rm Mpc}}\right),$$
where the fitted radius at 1 year is $(6.27\pm 0.22) \times 10^{16}$~cm
and the post-impact velocity is $v\spost = (4.12 \pm 0.22)\times
10^{16}$~cm~yr$^{-1}$, or $13\,040 \pm 690$~\kms.
The SSR of this fit was 3.7, and we plot the fitted function as the
blue line in Figure~\ref{fexpand}.

The SSR values for the powerlaw and the constant velocity fitted
functions were 2.4 and 3.7 respectively, and therefore our data do not
distinguish reliably between the two, although the powerlaw function
is a slightly better fit.  The values of SSR are close to the most
probable value for a Chi-squared distribution with 5 degrees of
freedom, $\chi^2_5 = 3$, indicating a reasonable fit, although we note
that our measurement errors are likely correlated, so the SSR is
likely not exactly $\chi^2$-distributed.
The slightly better fit of the powerlaw form may be due to the
constant-velocity period having ended and the powerlaw expansion
resuming, as is expected at late times after the impact of the ejecta
on the CSM shell \citep{HarrisN2020}.

\subsection{VLA Flux Density Measurements}
\label{svlaresults}

\begin{figure}
\centering
\includegraphics[width=0.48\textwidth]{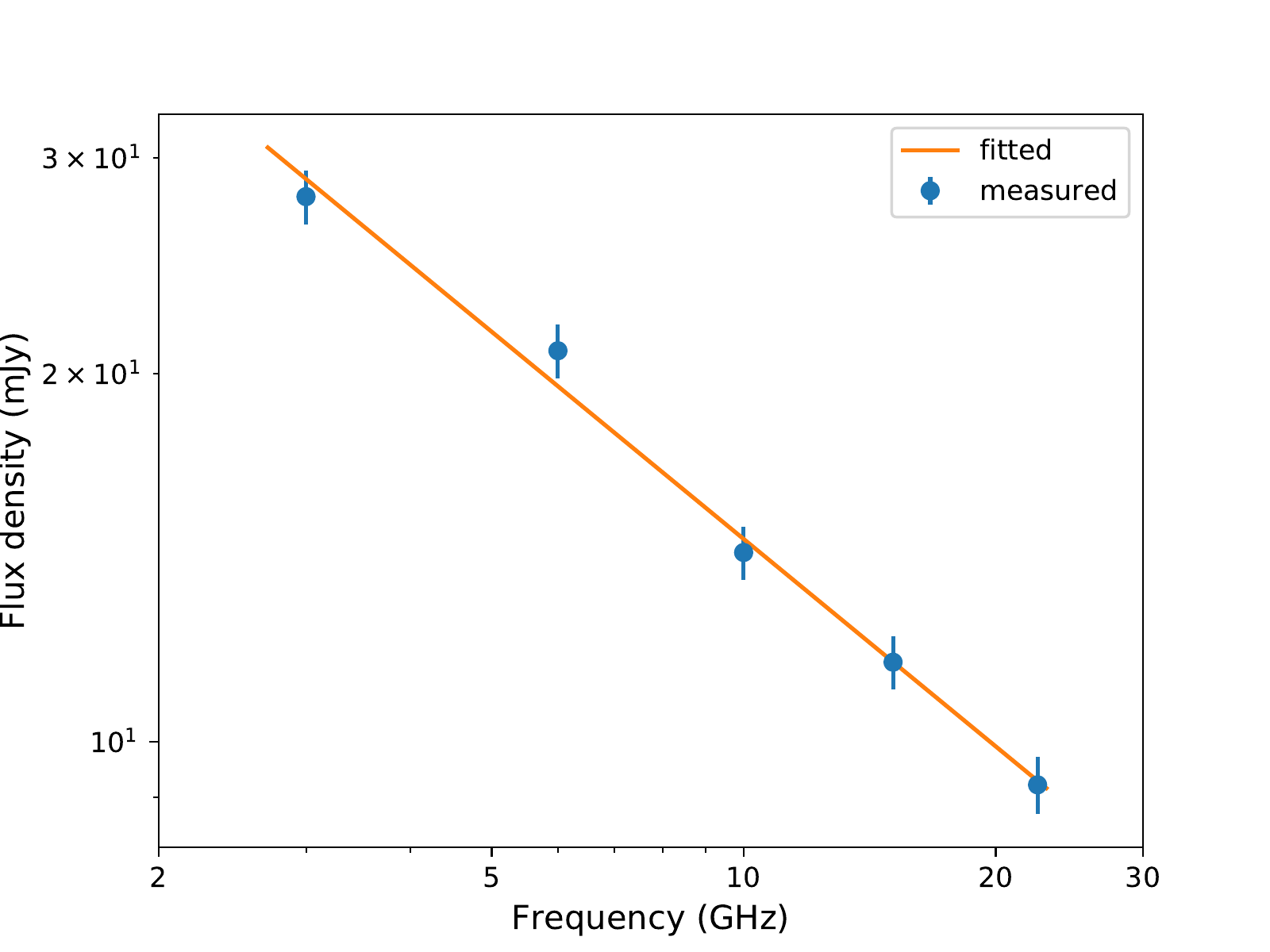}
\vspace{-0.1in}
\caption{The SED of SN~2014C on 2020 May 6, at $t = 6.3$~yr. The
  plotted errorbars show $1\sigma$ standard errors, which include both
  systematic and statistical contributions, and the straight line
  shows the powerlaw spectrum fitted to the data, which has a spectral
  index $\alpha = -0.56 \pm 0.03$.}
\label{fSED}
\end{figure}

On 2020 May 6, we measured the flux density of SN~2014C over a range of
frequencies between 3.0 and 23 GHz. We show the SED in
Figure~\ref{fSED}\@.  A powerlaw with spectral index,
$\alpha = -0.56 \pm 0.03$, (where $S_\nu \propto \nu^\alpha)$, and
$S_{\rm 5 \, GHz} = 21.6 \pm 0.6$~mJy fits all the measurements to
within the uncertainties, with the SSR (sum of squared residuals)
being 2.4, which is close to the expectation value of $\chi^2_3$.

\begin{figure}
\centering
\includegraphics[width=0.48\textwidth]{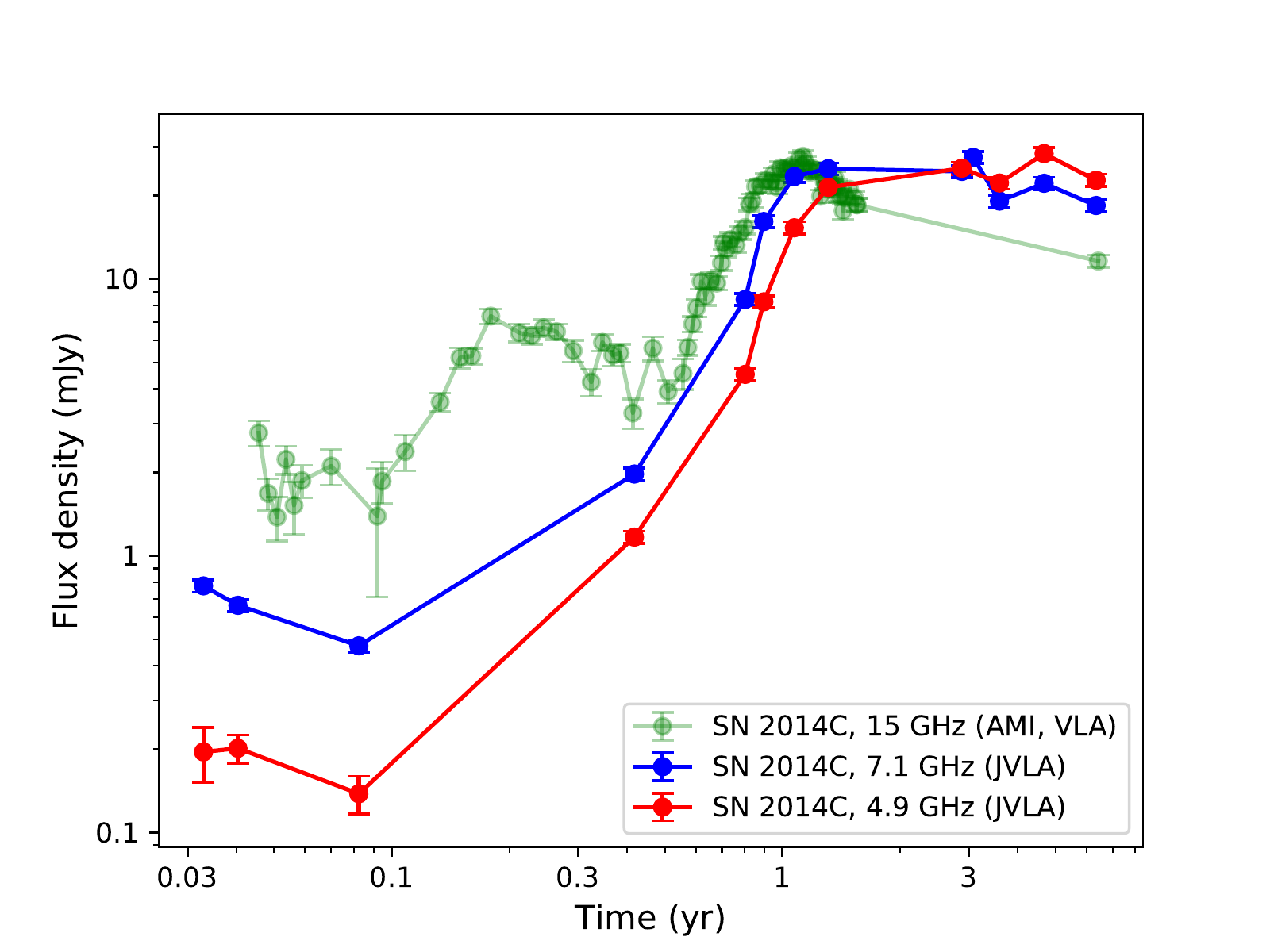}
\vspace{-0.1in}
\caption{The radio lightcurves of SN~2014C at three frequencies.  Red
  and blue show our VLA measurements at 4.9 and 7.1 GHz, respectively.
  Green shows the 15~GHz lightcurve, with all except the last point
  being at 15.7~GHz from AMI \citep{Anderson+2017}, and the last point
  being our own VLA measurement at 15.1~GHz (Table~\ref{tvla}).  We
  plot $1\sigma$ standard errors with statistical and dominating 5\%
  systematic contributions added in quadrature.  In many cases, the
  errorbars are smaller than the plotted points. The last pair of
  values at 4.9 and 7.1 GHz (at $t = 6.3$~yr) were interpolated
  between the measurements at 3.0 and 6.0 GHz, and 6.0 and 10.0 GHz
  respectively.}
\label{flightcurve}
\end{figure}

We give the flux densities measured from our VLA observations in
Table~\ref{tvla}.  We show the 4.9~GHz and 7.1~GHz lightcurves in
Figure~\ref{flightcurve}, where we also show for comparison the
15.7~GHz lightcurve measured by the Arcminute Microkelvin Imager from
\citet{Anderson+2017}.  The lightcurves do show the usual pattern of
an earlier rise at higher frequencies.  However, the overall nature of
the lightcurve is quite unusual, with a slow rise till $t \sim 0.6$~yr
that occurs in steps at least at 15 GHz, followed by a flat curve with
an almost value for the almost 6~yr since $t \sim 1$~yr.

\begin{table}
\begin{minipage}[t]{0.48\textwidth}
\caption{VLA Flux Density Measurements on 2020 May 6}
\label{tvla}
\begin{tabular}{c c}
  \hline
  \hline
Frequency (GHz)  & Flux Density\tablenotemark{a} (mJy) \\
 \phn 3.0  & $27.9 \pm 1.4$ \\
 \phn 6.0  & $20.9 \pm 1.1$ \\
  10.0     & $14.3 \pm 0.7$ \\
  15.1     & $11.6 \pm 0.6$ \\
  22.4     & $\phn 9.8 \pm 0.5$ \\
\hline
\end{tabular}
\\
\tablenotetext{a}{Our standard errors include the image background
  rms values and a 5\% flux-density calibration error, added in
  quadrature.}
\end{minipage}
\end{table}

\section{Discussion}
\label{sdiscuss}
\subsection{Morphology of SN 2014C}
\label{smorph}

The new VLBI image of SN~2014C at $t = 4.8$~yr (Fig.~\ref{fimage})
confirms the shell structure suggested by our earlier VLBI image from
$\sim$2~yr earlier \citepalias{SN2014C_VLBI}.  The source remains
relatively circular in outline.  Two enhancements in brightness are
visible to the east and west.  These are likely largely due to
convolution of a ring-like pattern with an elliptical restoring beam,
rather than being intrinsic brightness enhancements, fortuitously
aligned with the restoring beam. In our tests with synthetic data, a
completely circularly symmetric shell model produces an image very
similar to the observed one when convolved with our elliptical
restoring beam (see Section~\ref{svlbiimg}, Figure~\ref{fmodel}).

Although a spherical shell structure is consistent with our VLBI
image, could SN~2014C in fact have a different structure?
Given the observed image (Figure~\ref{fimage}), is it possible that
the source is intrinsically bipolar or elliptical, rather than having
a spherical shell structure?  Bi-polar jets occur in GRB and possibly
in some SNe \citep[e.g.][]{PapishS2011}.  SN~1987A, on the other hand,
has a structure that is axially, but not spherically symmetric
\citep[e.g.][]{McCrayF2016}.

To compare a possible bipolar structure to the spherical shell
structure, whose projection onto the sky-plane is circularly
symmetrical, we fitted a model consisting of two circular Gaussians to
directly the visibilities in the same way we fitted the spherical
shell model in Section~\ref{ssize}\@.  We found the fit considerably
poorer, despite the two-Gaussian-component model having more free
parameters.  We can therefore say that the observations disfavour a
simple bipolar structure.

A circular ring-like structure at some angle to the plane of the sky,
similar to SN~1987A's equatorial ring, is harder to constrain in this
manner, and if the ring is oriented near to the plane of the sky the
projected image will strongly resemble the projection of a thin
spherical shell.  A tilted ring structure of this nature could
therefore be also compatible with the VLBI image.  Future VLBI
observations at higher relative resolution may allow us to more
definitely determine the emission geometry.

\subsection{Radius and Expansion Speed}
\label{rradexp}

From our VLBI measurements, we determined the radius of the radio
emission region, which probably corresponds to the radius of the
forward shock \citep[see][for a discussion on the relationship between
the radio emission region and the forward shock in the case of
SN~1993J]{SN93J-4}.  At $t = 4.8$~yr, we measured a radius of
$\rx = 20.5 \pm 1.8$ (for $D = 15.1$~Mpc).  The velocity between our
previous measurement at $t = 2.9$~yr \citepalias{SN2014C_VLBI} and the
present one is $9400 \pm 2900$~\kms.

This velocity is consistent within the uncertainties, but lower by
$0.8\sigma$, than the value of $14\,000 \pm 4200$~\kms\ we found
between $t= 2.3$ and 2.9~yr \citepalias{SN2014C_VLBI},
suggesting that the shock front is likely decelerating somewhat.

We found in \S \ref{sexpand} that a powerlaw model fits all the VLBI
radius measurements marginally better than a constant velocity model
(see Fig.~\ref{fexpand}).  Our latest radius measurement ($t=4.8$~yr)
suggests a possible increase in the deceleration at the latest times,
in either the constant-velocity or the powerlaw models of the
expansion.  Further VLBI measurements should be undertaken to better
constrain any change in deceleration.

Given the complicated nature of SN~2014C's CSM, with,
going outward, first a low-density cavity, then a very dense shell,
then a moderately dense stellar wind, the real expansion curve will be
more complex than a simple powerlaw.  In \citetalias{SN2014C_VLBI} we
compared the evolution of SN~2014C to scaled hydrodynamic simulations
from \citet{vMarle+2010} to show the generally expected behaviour of a
SN shock slowing down dramatically upon first encountering a thick
shell, but the shock speed then recovering somewhat.

Since then, \citet{HarrisN2020} have performed new hydrodynamic
simulations more specifically aimed at cases like SN~2014C\@.  They
find the measurements can be accounted for with the following model:
SN~2014C explodes in a low density cavity.  The ejecta first impact on
the dense CSM shell at $t \sim 0.3$~yr (100~d), then exit the dense
shell again at $t \sim 0.5$~yr (190~d), subsequently interact with a
wind medium with $\rho \propto r^{-2}$ corresponding to a period
before the ejection of the dense shell, where the progenitor was
loosing mass relatively steadily.  This model is also consistent with
the VLBI radius measurements, however, as we show in
\S~\ref{slightcurve} below, the radio lightcurves suggest a slight
variation.

\subsection{Spectral Energy Distribution}
\label{sSED}

We found that the spectrum of the radio emission at $t= 6.3$~yr was
well described by a powerlaw spectrum, with
$S_\nu = (21.6 \pm 0.6 \; {\rm mJy}) \cdot (\nu/(5\, {\rm GHz})
^\alpha$ with $\alpha = -0.56 \pm 0.03$.  Such a spectrum is what is
expected from optically-thin emission resulting from the SN shock,
although $\alpha$ is somewhat flatter than usual: \citet{Weiler+2002}
fitted the optically-thin values of $\alpha$ for 14 different SNe, and
our value for SN~2014C is close to their flattest value of
$\alpha = -0.55$ (which was for the Type IIL SN~1970G).

The relatively flat spectrum of SN~2014C might just be due to a slow
transition from optically-thick (inverted spectrum) to optically-thin.
As can be seen from the 4.9 and 7.1~GHz lightcurves in
Fig~\ref{flightcurve}, the spectrum between these two frequencies
remained inverted until $t \sim 3$~yr.  If the spectrum were still
transitioning between optically thick and thin, one would expect
significant curvature in the spectrum, with a steep spectrum at high
frequencies and a flat (or inverted one) at low frequencies.  Indeed,
our lowest-frequency measurement at 3 GHz suggests a marginally
flatter spectrum below 6 GHz.  If we fit only the points at 6 GHz and
above, we obtain $\alpha = -0.61 \pm 0.04$, which is within the normal
range.

\citet{SN93J-2} found also that the optically-thin value of $\alpha$
for the Type IIb SN~1993J became flatter with time.
\citet{Maeda2013a} shows that such a flattening is in fact expected,
with the shock acceleration being less efficient in young SNe where
the shock speed is high, leading to steeper spectra, and becoming more
efficient later on, leading to flatter spectra for supernova remnants,
for which $\alpha$ clusters around the expected value for shock
acceleration of $\alpha = -0.5$.  This process may also be occurring in
SN~2014C, and contributing to the relatively flat value of $\alpha$.

\subsection{Radio Lightcurve}
\label{slightcurve}

From our VLA observations at $t = 6.3$~yr we found that the lightcurve
has an extended, almost constant, plateau since $t \sim 0.8$~yr.  Such
a lightcurve is unusual, the lightcurves of the majority of SNe show
an approximately powerlaw decline after a time on the order of one
month \citep[e.g.][]{Weiler+2002, RadioLumFn}.

The strong and sustained radio emission is interpreted as being due to
the strong CSM interaction as the forward shock ploughed through the dense
CSM shell.  The rate of particle acceleration is dependent on the CSM
density but is also strongly dependent on the shock velocity.
\citet{HarrisN2020} show that while the CSM density drops when the
shock emerges from the dense shell, the shock speed increases which
can lead to an increase in radio emission despite the drop in CSM
density.

However, these increases in the shock velocity are temporary, and since
the shock emerged from the dense shell some time ago, why is the radio
brightness still staying high? 
In the self-similar mini-shell model of a SN, where both
the ejecta and the CSM density structures are powerlaws in radius, the
radio brightness evolves as $S \propto t^\beta$
\citep{FranssonLC1996}, with
$$ \beta = -\{3 - \alpha - [6 - \alpha - (s/2)(3 - \alpha)][(n-3)/(n-s)]\},$$
where the density of the CSM is $\propto r^{-s}$, that of the ejecta
is $\propto r^{-n}$, and $\alpha$ is the radio spectral index.
Although in the case of SN~2014C, it is clear that the CSM structure
is more complex than a simple powerlaw, and strictly self-similar
evolution is therefore not expected, it is nonetheless instructive to
compare SN~2014C's evolution to expectations from the self-similar
case.  It is expected that at some point after the shock has passed
through the dense CSM, the evolution would once again approach being
self-similar.  Since the radio brightness of SN~2014C has not
declined much, $\beta \simeq 0$\@.  For a typical value of $n=16$ and
our observed $\alpha = -0.56$, in a self-similar scenario we would
have $s = 1.45$.  Since SN~2014C's evolution was not self-similar, the
actual value of $s$ will probably differ. However, the shock exited
the dense CSM shell already at $t \sim 0.3$~yr, so by $t\sim 6.3$~yr
probably SN~2014C's evolution is again approaching the self-similar
solution, and that value of $s$ at least approximately correct.  The
flat lightcurve therefore probably suggests a CSM density profile
notably flatter than that for a steady wind (density
$\propto r^{-2}$)\@.  The exact value of $n$ has only a minor effect
on this conclusion, which holds for any reasonable value of $n$.

\citet{HarrisN2020} suggest that SN~2014C's shock crossed through a
dense shell of CSM, and is now interacting with a wind CSM, with
$s = 2$.  While this scenario fits the measured sizes and expansion
velocities, it is hard to reconcile with the lack of any decay in the
radio luminosity.  The flat lightcurve suggests $s \sim 1.5$, implying
that the shock is currently interacting with CSM from a period where
the progenitor's mass-loss was relatively steadily decreasing with
time.  \citet{HarrisN2020} found that both models with $s=2$ (a steady
wind) and $s = 1$ (wind with density decreasing with time) were
compatible with the measurements, therefore a model with $s = 1.5$
should also be compatible with the data.

The decrease in time of the mass-loss rate of the progenitor must have
occurred only over a bounded period, and there was likely steadier
mass-loss rate before the decrease.  The shock radius is currently
$2.05 \times 10^{17}$~cm.  If we assume a wind speed of 1000~\kms,
typical of a Wolf-Rayet like progenitors,
the shock is currently interacting with material lost from the star
only about a century before the explosion.  Even if the wind speed was
10~\kms, typical of supergiants, the age of the material is only of
order $10^4$~yr.  Fluctuations in the mass-loss rate over these
timescales, short compared to the age of the star, have been seen or
inferred in a number of stars \citep[e.g.][]{Smith2014}.
It is likely, therefore, that the lightcurve will turn over and
decrease in the future as the shock moves beyond the region formed by
the mass-loss that was declining in time prior to the massive shell
ejection shortly before the explosion.

\section{Summary and conclusions}
\label{ssummary}

We report on our new VLBI and VLA observations of SN~2014C\@.  We
resolved the radio emission from the expanding shell of ejecta and
determined the radius of the emission region at $t = 4.8$~yr after the
explosion.  Comparing these results with those of our earlier VLBI
measurements, we found the following:

\begin{trivlist}

\item{1.} Our new VLBI observations show a structure which is
  relatively circular in outline and enhanced towards the outer edge.
  There is a clear enhancement to the east and west, much of which is
  likely not intrinsic, but rather due to the convolution with an
  elongated restoring beam.  Some intrinsic enhancement of the surface
  brightness does however seem likely, particularly to the east.  The
  observed image is compatible with a relatively thin spherical shell
  seen in projection.  Our model fits show that a simple bipolar
  structure is unlikely.  A ring-like structure, however, could also
  be compatible with the image.
  
\item{2.} At $t = 4.8$~yr, the angular outer radius of the supernova was
  $0.91 \pm 0.08$~mas, corresponding to
  $(20.5 \pm 1.8) \times 10^{16}$~cm (for a distance of 15.1~Mpc). The
  speed between the last two epochs of VLBI observations ($t = 2.9$
  and 4.8~yr) was $9400 \pm 2900$~\kms.

\item{3.} The expansion of SN~2014C, as determined from VLBI
  observations ($t = 1.1 - 4.8$~yr) is compatible with powerlaw
  expansion, with $r \propto t^{0.77 \pm 0.03}$, suggesting a moderate
  amount of deceleration over the SN's lifetime.  The measurements are
  compatible with an early deceleration, and an approximately
  constant-velocity expansion with $13\,040 \pm 690$~\kms\ since
  $t = 1.1$~yr.  There is a suggestion that the deceleration is
  increasing again after $t \sim 3$~yr.

\item{4.} The radio spectral energy distribution is consistent with a
  powerlaw with $S \propto \nu^\alpha$ where
  $\alpha = -0.56 \pm 0.03$.  This value of $\alpha$ is somewhat
  flatter than that seen in the majority of SNe.  There is a hint of
  flattening of the spectrum below 6~GHz, as might be expected if it
  were just now becoming optically thin at low frequencies.
  
\item{5.} The radio lightcurve at $\sim$6~GHz had reached a peak of
  $\sim$25~mJy, corresponding to a $\nu L_\nu$ luminosity of
  $4.1 \times 10^{37}$~erg~s$^{-1}$, after about one year, and has
  stayed almost constant since then, up to our latest measurement at
  $t = 6.3$~yr.

\item{6.} Our observations are consistent with a picture that has
  emerged of SN~2014C having a mass-loss event that ejected a very
  dense shell not long before the explosion, with the mass-loss rate
  prior to the shell ejection being much lower.

\item{7.} The sustained radio emission since $t \sim 1$~yr suggests
  that the progenitor went through a period of steadily decreasing
  mass-loss before ejecting the dense shell and then exploding as a SN.

\end{trivlist}

\section*{Acknowledgements }

We thank the teams of both the EVN and the VLA for their work to make
the observations possible.  The EVN is a joint facility of independent
European, African, Asian, and North American radio astronomy
institutes.  The National Radio Astronomy Observatory is a facility of
the National Science Foundation operated under cooperative agreement
by Associated Universities, Inc.  We have made use of NASA's
Astrophysics Data System Abstract Service. This research was supported
by both the National Sciences and Engineering Research Council of
Canada and the National Research Foundation of South Africa.

We also thank the anonymous referee for his or her comments
which improved the paper.

\section*{Data Availability Statement}

The raw data underlying this paper are available in EVN and NRAO
archives, and can found under the observing codes EB066A, EB066B for
the EVN and 20A-441 for NRAO\@.  The calibrated data or images
underlying this paper will be shared on reasonable request to the
corresponding author.

\bibliographystyle{mnras}
\bibliography{mybib1,sn2014c_temp,sn2014c_temp2}

\label{lastpage}

\clearpage

\end{document}